\documentclass[reprint,twocolumn
fleqn, amsmath,amssymb, aps,
pra, prb, rmp, prstab, prstper, floatfix
]{revtex4-1} 
\usepackage{color} 
\usepackage{graphicx}
\usepackage{dcolumn}
\usepackage{tikz,xcolor,hyperref}
\usepackage[utf8]{inputenc}
\DeclareUnicodeCharacter{2212}{-}

\definecolor{lime}{HTML}{A6CE39}
\DeclareRobustCommand{\orcidicon}{%
        \begin{tikzpicture}
        \draw[lime, fill=lime] (0,0)
        circle [radius=0.16]
        node[white] {{\fontfamily{qag}\selectfont \tiny ID}};
        \draw[white, fill=white] (-0.0625,0.095)
        circle [radius=0.007];
        \end{tikzpicture}
        \hspace{-2mm}
}
\usepackage[mathlines]{lineno}
\foreach \x in {A, ..., Z}{%
\expandafter\xdef\csname orcid\x\endcsname{\noexpand\href{https://orcid.org/\csname orcidauthor\x\endcsname}
{\noexpand\orcidicon}}
}

\begin{document}

\title{Chiral edge transport along domain walls in magnetic topological insulator nanoribbons}

\newcommand{\KTH}{Department of Applied Physics, School of Engineering Sciences, KTH Royal Institute of Technology, 
AlbaNova University Center, SE-10691 Stockholm, Sweden}
\newcommand{\LNU}{Department of Physics and Electrical Engineering, Linn{\ae}us University, 391 82 Kalmar, Sweden}
\author{N. Pournaghavi\orcidA}   \email{nezhat@kth.se (Corresponding author)}
    \affiliation{\KTH} 
\author{C.~M.~Canali\orcidB}   
    \affiliation{\LNU}

\begin{abstract}
Quantum anomalous Hall insulators are topologically characterized by non-zero integer Chern numbers, the sign of which depends on the direction of the exchange field that breaks time-reversal symmetry. This feature allows the manipulation of the conducting chiral edge states present at the interface of two magnetic domains with opposite magnetization and opposite Chern numbers.
Motivated by this broad understanding, the present study investigates the quantum transport properties of a magnetized $Bi_2Se_3$ topological insulator nanoribbon with a domain wall oriented either parallel or perpendicular to the transport direction. Employing an atomistic tight-binding model and a non-equilibrium Green's function formalism, we calculate the quantum conductance and explore the nature of the edge states. We elucidate the conditions leading to exact conductance quantization and identify the origin of deviations from this behavior.
Our analysis shows that although the conductance is quantized in the presence of the horizontal domain wall, the quantization is absent in the perpendicular domain wall case. Furthermore, the investigation of the spin character of the edge modes confirms that the conductance in the horizontal domain wall configuration is spin polarized. This finding underscores the potential of our system as a simple three dimensional spin-filter device.
\end{abstract}

\maketitle

\section{Introduction}
Topological materials are a new state of matter that display topologically protected conducting boundary states, e.g., at the edges of a two-dimensional (2D) sample or on the surface of a three-dimensional (3D) system\cite{Hsieh2009,hasan2010, XLQi, Ando2013,Bansil2016}. The existence of these boundary states is ensured by non-zero topological indexes or invariants of the bulk band structure, an interplay usually refereed as boundary-bulk correspondence. It is precisely at the interface between two regions with different bulk topological invariants (one possibly being the vacuum) that topological boundary states emerge. The integer quantum Hall effect in a 2D electron gas in the presence of a strong perpendicular magnetic field discovered more than 40 years ago\cite{Klitzing1980} was the first example of this correspondence. Here the 2D bulk topology is characterized by a non-zero Chern number  equal to the quantized Hall conductance and to the number of chiral gapless edge states responsible for the dissipationless transport. In more recent years, topological band theory has found that 2D and 3D topological insulators (TI) display similar gapless boundary states related to another bulk topological invariant, the $Z_2$ number, which is protected most notably by time-reversal symmetry (TRS)\cite{Hsieh2009,hasan2010, XLQi} (or by a crystal symmetry in case of topological crystal insulators).  When TRS is broken by the presence of magnetism either in the bulk (via magnetic doping\cite{Checkelsky2012,sessi2016, Islam2018}) or on the surfaces of a TI thin film (via surface doping or by proximity to a magnetic layers\cite{Vobornik2011}) the 2D surface states acquire an energy gap, signalling a topological phase transition to either a Chern or to an axion insulator\cite{DVbook}. In the first case, when the film is arranged into a Hall-bar geometry, a quantum transport phenomenon known as 
Quantum Anomalous Hall Effect (QAHE) emerges\cite{He2013,Chang_Li2016,AHM_RMP_2023}. 
Predicted in 2013\cite{Yu2010}, the QAHE has been observed in uniformly-doped magnetic TI thin films\cite{chang2013},
in modulation-doped TI films \cite{mogi_APL2015, mogi_NatMat2017, mogi_ScAdv2017,xiao2018, Allen2019, Mogi2022} that
have surface magnetism only, and recently also in TI films with proximity-induced 2D magnetism\cite{mogi_APL2019}.
The QAHE is a quantum Hall effect without Landau levels\cite{haldane1988} due to spontaneously broken TRS. It is characterized again by the same topological bulk integer Chern number of the ordinary QHE, which is equal to number of spin-polarized gapless chiral edge states hosted on the sidewalls of the TI film and carrying a dissipationless current. 

The QAHE dissipationless chiral edge states have a great potential in spintronics for the next generation of information processing devices operating at zero magnetic field with low power 
consumption\cite{tokura_NatRev2019, QL_HE_2022}. 
Furthermore, it is possible to envision the realization of more complex topological chiral networks based on 1D chiral interface states which, according to topological band theory, also appear at the boundary of two QAHE insulators with different Chern numbers 
$\cal C$\cite{tokura_NatRev2019}. The difference 
in $\cal C$ between two adjacent QAHE insulator domains determine the number of chiral interface states at a given edge. 

One simple way to generate such chiral interface states is to create two regions of opposite magnetizations in a QAHE insulator, separated by a sharp magnetic domain wall (DW). The experimental realization of these systems has shown evidence of quantized chiral edge transport\cite{Yasuda2017, Rosen2017}, but has been so far quite challenging.
However, quite recently, by employing an in-situ mechanical mask in magnetically doped $(Bi,Sb)_2Se_3$ multilayers, Zhao et, al. have used molecular beam epitaxy to synthesize efficiently robust QAHE 1D junctions along a DW separating two regions of opposite magnetization with Chern number $\cal \pm C$. In a QAHE bar with a DW orthogonal to the current direction, they demonstrated the existence of two parallel chiral edge states along the DW leading to quantized transport at zero magnetic field\cite{Zhao2023}.

Quasi-1D chiral edge and interface states and their quantum transport properties in QAHE set-ups in the presence of DWs have been studied theoretically in a few papers using effective continuum models for the magnetic TI surfaces ~\cite{Wakatsuki2015,Sedlmayr2020,Han2023}. The nature of the states at the domain wall can be controlled by an external electric field. This is due to the fact that depending on the position of the Fermi energy, the chirality as well as the coupling between the magnetic moments with TI states can change~\cite{Wakatsuki2015}.
It can be shown that the equilibrium charge current along the domain wall in a 3D TI in the presence of a local Zeeman field, is equal to the sum of counter-propagating equilibrium currents flowing along the external boundaries of the domains~\cite{Sedlmayr2020}. The number of chiral interface modes along the DW is determined by the difference of Chern numbers of adjacent regions, the tuning of which can be used to manipulate the current partition 
at the DW junctions\cite{Han2023}. Note that earlier theoretical work had predicted that robust edge and interface states should occur in graphene heterostructure QHE bars along the edges and the DWs created at the interface of two adjacent regions under the effect of oppositely oriented magnetic fields. In this case, when the DW is orthogonal to the direction of the current, the DW interface states affect the edge states by opening a back scattering channel which prevents the perfect quantization of the two-terminal conductance~\cite{Lado2013}.

In this paper, we consider a nanoribbon of magnetized $Bi_2Se_3$ 3D TI, hosting a magnetic domain wall (DWs) either parallel or perpendicular to the nanoribbon direction, see Fig.~\ref{struct}. The purpose of the study is to investigate how spin-polarized quasi 1D chiral states along the interface of two magnetic regions with opposite bulk topological invariants, affect the electronic and spin transport. In particular, we explore the influence of these interface states and possible bulk states on the quantized conductance carried by the chiral edge states of the nanoribbon, which are responsible for the QAHE in a system without DWs. We use an atomistic tight-binding model\cite{Pertsova2014, pertsova2016} which provides an adequate microscopic description of the electronic band structure, along with the non-equilibrium Green's function formalism to study the two-terminal longitudinal conductance. Importantly, in contrast with the case of effective models employed in previous studies, in our approach not only dissipationless 1D chiral edge/interface states but also dissipative bulk and non-chiral side-wall states, which are always present in the system~\cite{pertsova2016, Pournaghavi2021, Pournaghavi2021_et_al_2021}, emerge directly from the band structure and are treated on the same footing.
The use of a microscopic tight-binding model adds considerable computational complexity but it provides an unbiased description of physics and realistic estimate of the typical system sizes where the effects described here are likely to be observed. On the basis of the spin character of the edge states we also propose to use this set-up as a spin filter device. 

The paper is organized as follows. In Sec.II we describe the theoretical tight-binding model and the non-equilibrium Green's function approach to quantum transport. Sec. III presents the results of the electronic structure and conductance. Sec. IV contains concluding remarks.

\section{\label{sec:level2}THEORETICAL MODEL AND TRANSPORT FORMALISM}

\subsection{\label{sec:level21} Tight-binding model for a thin-film heterostructure nanoribbon}

\begin{figure}[b] \centering
\includegraphics[width=7.5cm,height=9cm,keepaspectratio]{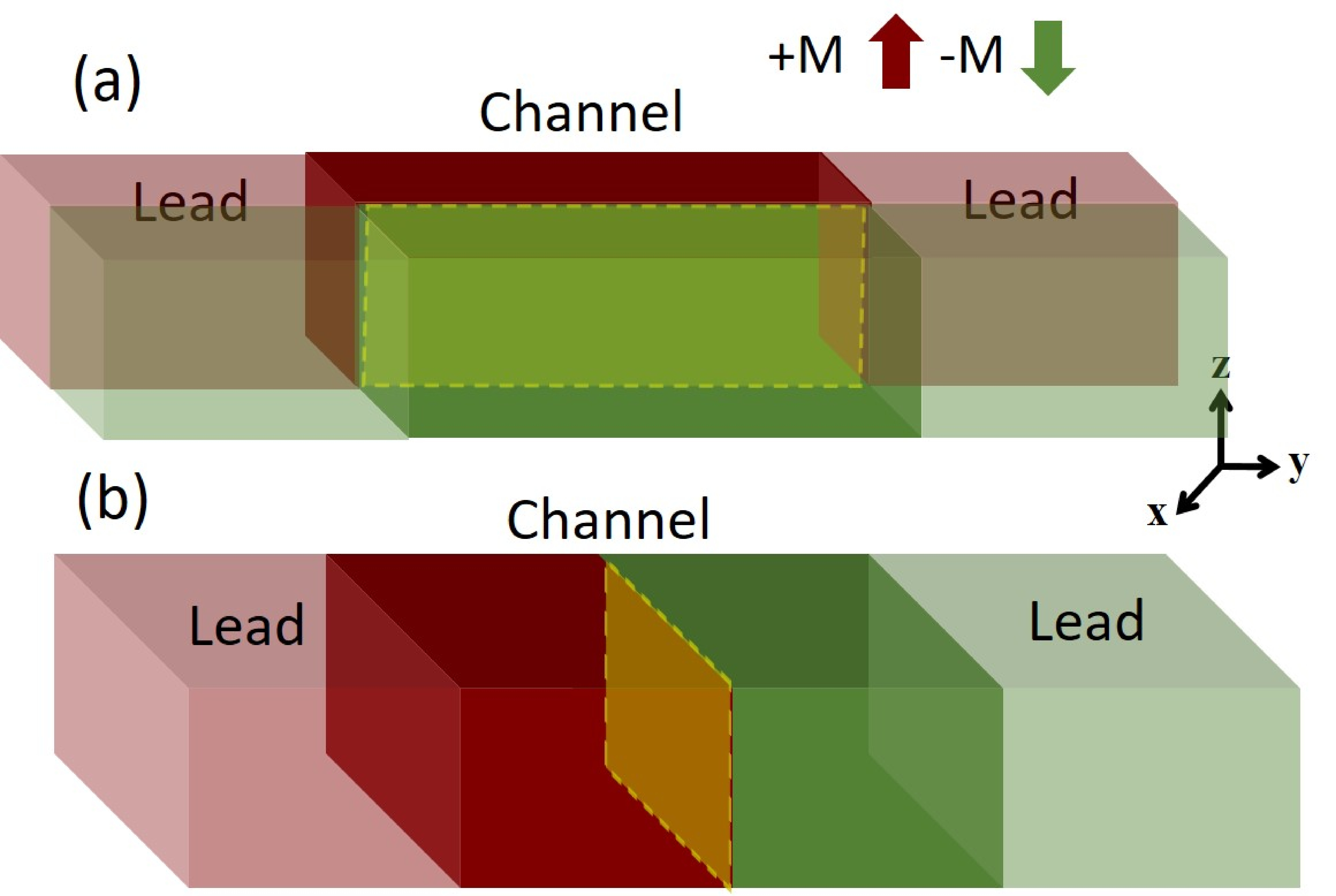}
\caption{\label{struct} {Schematic picture of a magnetized TI nanoribbon attached to two semi-infinite leads from left and right with (a) a parallel domain wall (b) a perpendicular domain wall. The yellow shaded area indicates the domain wall.
}} \end{figure}

In order to consider a domain wall in the system, we study two distinct two-terminal heterostructures with a parallel and a perpendicular domain wall as shown in Fig.~\ref{struct}. In both cases a homogeneous exchange field is considered all over the ribbon as a result of a magnetic substrate with a high Curie temperature which induces a proximity coupling to the nearby layers of the $Bi_2Se_3$. 
The direction of the magnetization however depends on the two cases that we have considered. We make the simplifying assumption that the magnetization changes abruptly at the interface since we are aiming at developing a qualitative analysis and clarify the basic physical concepts rather than attaining precise numerical results. However, it's important to note that earlier work~\cite{Lado2013} on transport in graphene in the presence of DWs has shown that a linearly smooth transition introduces just some small fluctuations in the conductance steps; yet the qualitative outcome remains consistent with that one obtained assuming abrupt DWs.  Also, the recent  results shown in Ref \cite{Zhao2023} point to DWs that are indeed rather narrow (of the order of a few nm). Additionally, atomically sharp DWs in an antiferromagnet thin film have been realized recently \cite{Krizek2022}.

To model $Bi_2Se_3$ thin films we use an atomistic tight-binding (TB) model based on s and p orbitals of Bi and Se atoms ($sp^3$ )\cite{Pertsova2014, pertsova2016}. TB model incorporates atomic orbitals and their interactions, providing a microscopic description of electron dynamics. By considering the spin-orbit coupling and crystal symmetry, this model effectively describes the emergence of topological surface states. The Hamiltonian can be written as follows 
\cite{Kobayashi2011,Pertsova2014, pertsova2016}

\begin{equation}\label{eq:2}
\begin{aligned}
H_C &= \sum _{ii',\sigma\alpha\alpha'}t_{ii'}^{\alpha\alpha'}e^{i{\bf k}\cdot {\bf r}_{ii'}}c_{i\alpha}^{\sigma\dag}c_{i'\alpha'}^{\sigma}\\ 
  &\quad + \sum _{i,\sigma\sigma',\alpha\alpha'} \lambda_i <i,\alpha,\sigma|{\bf L} \cdot {\bf S}| i,\alpha',\sigma'> c_{i\alpha}^{\sigma\dag}c_{i\alpha'}^{\sigma'}\\
  &\quad +   \sum _{i,\sigma ,\alpha} M_i c_{i\alpha}^{\sigma\dag }
{\sigma}_{z}^{\sigma \sigma}c_{i\alpha}^{\sigma\dag}\;.
\end{aligned}
\end{equation}

 The first (kinetic-energy) term of Eq.~\ref{eq:2} contains Slater$-$Koster hopping and site-energy parameters\cite{Slater1954}, which in the present case have been extracted from density functional theory (DFT) calculations\cite{Kobayashi2011}. Here $c_{i\alpha}^{\sigma\dag}(c_{i\alpha}^{\sigma})$ is the creation (annihilation) operator for an electron with spin $\sigma$ and atomic orbital $\alpha \in$ ({$s, p_x, p_y p_z$}) at site i. {\bf k} is the reciprocal-lattice vector that spans the Brillouin zone. $i^\prime \neq i$ runs over all neighbors of atom i in the same atomic layer as well as the first and second nearest-neighbor layers in the adjacent cells, and {\bf $r_{ii\prime}$} represents the vector connecting two neighbor atoms.
In the second term, an on-site spin-orbit coupling (SOC) interaction is implemented in the intra-atomic matrix elements \cite{walteraharrison1999}, in which $|i,\alpha,\sigma>$ are spin- and orbital-resolved atomic orbitals. ${\bf L}$ and ${\bf S}$ are the orbital angular momentum and the spin operators, respectively, and $\lambda_i$ is the SOC strength \cite{Kobayashi2011}.  The last term represents the exchange field with the strength $|M_i|=0.2\; eV$ at different sites and can be either positive or negative with respect to the positive direction of the 
$z$-axis\cite{Pertsova2014, pertsova2016}. This is the crucial term for breaking TRS in the system, which opens an exchange band gap at the Dirac point of the surface states and induces a finite Berry curvature. Below we will specify the exchange field configuration for two different cases.

\subsection{\label{sec:level22} Green's function and transport formalism}
In order to calculate the transport properties, we use the non-equilibrium Green's function (NEGF) method which describes the propagation of electrons in a system under the influence of an external voltage or perturbation. Given the Hamiltonian of the central channel at the $\Gamma$ point, 
the spin-dependent retarded ($r$) and advanced ($a$) Green\noindent 's functions are given by \cite{0521599431}

\begin{equation}\label{eq:3}
\begin{aligned}
\boldsymbol{{\cal G}}^{r} (E)&=(E^{+}\boldsymbol{I}-\boldsymbol{H}_C - \sum_n \Sigma_n(E))^{-1}\\
\quad&=\left[{\boldsymbol{\cal G}}^{a}(E)\right]^{\dag },
\end{aligned}
\end{equation}
where $E^+\equiv E + i 0^+$ and $\boldsymbol{I}$ is the identity matrix. $H_C$ is defined in Eq~\ref{eq:2} in which $k=0$. The sum $n= L, R$ is over the two self-energies, which account for the interactions and boundary conditions due to the connection of the left (right) semi-infinite electrode to the central channel.
The self-energies can be expresses as
\begin{equation}\label{eq:4}
\begin{aligned}
\boldsymbol{\Sigma} _{L(R)}(E)= \boldsymbol{H}_{L(R),C}^{\dag }  \boldsymbol{g}_{L(R)}(E)\boldsymbol{H}_{L(R),C},
\end{aligned}
\end{equation}
where $\boldsymbol{H}_{L(R),C}$ is the tunneling Hamiltonian between the central region and the left(right) lead. The Green's function $\boldsymbol{g}_{L(R)}$ is the {\it surface} Green's function of the left(right) lead that captures the specific behavior of electrons at the material's surface and can be calculated using the Sancho-Lopez-Rubio recursive method~\cite{Sancho1984}. In this method, the semi-infinite lead is divided into layers perpendicular to the interface and each layer is treated as a finite system, allowing for the calculation of its Green's function. This recursive method relates the Green's function of a layer to the Green's function of the neighboring layer through the self-energy term. We first calculate the effect of the surface connected to the right semi-infinite lead  using this iterative method ~\cite{PhysRevB.60.7828}, which is then used to connect the central region to the leads in such a way that a unit cell from the central region is added to the right lead one at a time. In this way, a surface Green’s function $\boldsymbol{g}_{R}^{l}$ is calculated for each unit cell $l$ starting from $l=M$, where M is the number of unit cells in the central region, to $l=2$. It should be noted that the surface Green's function is calculated using Hamiltonian of the leads , which may differ from that of the central region. However, here we are considering the leads to be simply an extension of the central part which therefore do not introduce any additional scattering effects, focusing solely on the influence of the the DWs. Additionally, due to ballistic transport, the length of the channel does not impact the results. The $\boldsymbol{g}_{R}^{l}$s are calculated from $\boldsymbol{g}_{R}^{l+1}$ according to

\begin{equation}\label{eq:5}
\boldsymbol{g}_{R}^{l}=(E^{+}\boldsymbol{I}-\boldsymbol{H}_C - \boldsymbol{H}_{R,C}\boldsymbol{g}_{R}^{l+1}\boldsymbol{H}_{R,C}^{\dag })^{-1}\;.
\end{equation}
Once we have obtained the Green's function of the whole system, we can calculate the longitudinal two-terminal conductance as

 \begin{equation}\label{eq:7}
 \begin{aligned}
G (E)&=\frac{e^{2} }{h}
\mathrm{Tr} \left[ \boldsymbol{\Gamma} ^{L} (E) \boldsymbol{\cal G}^{r} (E) \boldsymbol{\Gamma}^{R} (E) \boldsymbol{\cal G}^{a} (E)\right]\\
\quad&=\frac{e^{2} }{h} T(E),
 \end{aligned}
\end{equation}
where
$\boldsymbol{\Gamma}^{L/R} = i[ \boldsymbol{\Sigma}^r_{L/R}- (
\boldsymbol{\Sigma}^r_{L/R} )^\dag]$ is the broadening due to the left and right electrode contacts.

In order to calculate the steady-state local transport properties in the full
NEGF formalism, we need both the retarded and lesser Green\noindent 's
functions, which contain information about the density of available states and
how electrons occupy these states, respectively. In the phase-coherent regime
where interaction self-energy functionals are zero, by using the Keldysh
formalism we can simply write the lesser Green\noindent 's function in terms of
the retarded Green\noindent 's function and the broadening matrices as
\citep{Cresti2003}
\begin{equation}\label{eq:8}
 \begin{aligned} &&{\cal G}_{i \sigma , j \sigma'}^< (E)
=[\boldsymbol{\cal G}^r (E) \boldsymbol{\Sigma}^{<}_L+\boldsymbol{\Sigma}^{<}_R
\boldsymbol{\cal G}^a (E)]_{i \sigma, j \sigma'} \nonumber \\
&&=i[\boldsymbol{\cal G}^r(E)(f_L(E)\boldsymbol{\Gamma}^{L} + f_R (E)
\boldsymbol{\Gamma}^{R}) \boldsymbol{\cal G}^a(E) ]_{i \sigma , j \sigma'} .
\quad \quad  \end{aligned}
\end{equation}
Once the lesser Green's function is known, we can also calculate both the
equilibrium and the out-of-equilibrium (when an applied voltage is present), spin-resolved, local density of states $n_i^\sigma (E)$ at space position $i$ and at a given energy $E$
\citep{Nikolic2006, Chen2010}
\begin{equation}\label{eq:9} n_i^\sigma (E) =\frac{e}{4\pi} {\cal G}_{i\sigma,i \sigma}^< (E)\, .
\end{equation}

\section{\label{sec:level3}RESULTS AND DISCUSSION}

\subsection{\label{sec:level31} Parallel domain wall}

\begin{figure}[bh] \centering
\includegraphics[width=8.5cm,height=10cm,keepaspectratio]{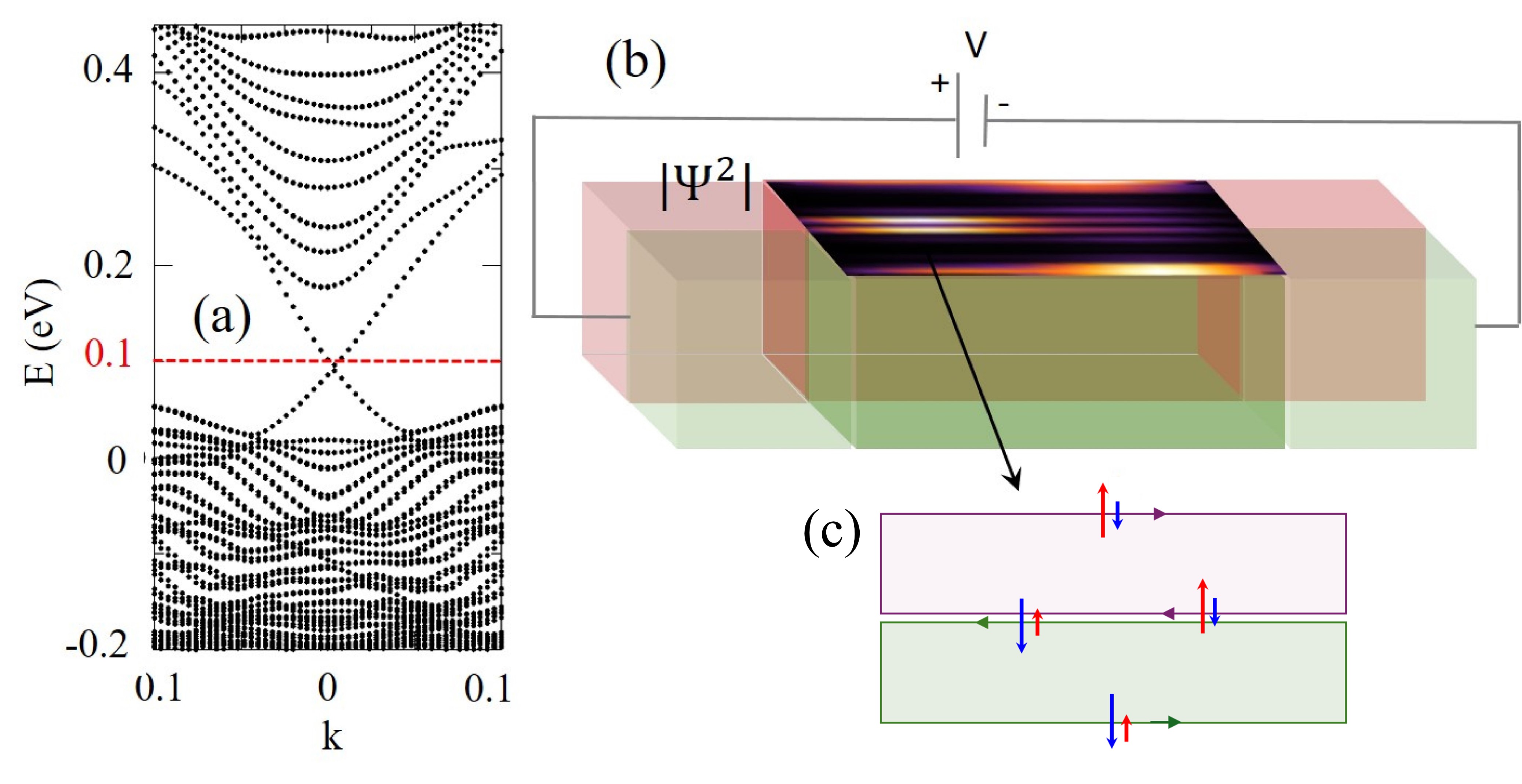}
\caption{\label{WF} {(a) Quasi-1D band structure of an infinitely long nanoribbon of magnetized Bi$_2$Se$_3$ TI of x-width 21 nm and z-thickness 2.6 nm, with a DW along the longitudinal $y$-direction. 
The thickness of the ribbon corresponds to 3 quintuple layers (b) Spatial distribution of the modulus square of the wave function $|\Psi|^2$ for the two chiral right-moving edge states (localized at the external edges of the nanoribbon) and the two chiral left-moving states localized along the DW in the middle of the nanoribbon,  all at quasi-degenerate energy $E=0.1eV$, plotted at the top surface. The spin character of the edge and interface states is also shown schematically in panel (c) featuring the top-view of the nanoribbon surface (pointed by the black arrow), where red (blue) arrows indicate spin-up (-down) states. 
}} 
\end{figure}

We start by considering the heterostructure in Fig~\ref{struct}(a) (parallel domain wall along the longitudinal direction $y$). The system is translational invariant along the $y$-direction, and we can plot the quasi-1D band structure as a function of the wave vector $k_y$. Since we have a nanoribbon, both the width and the thickness of the central channel should be chosen carefully to avoid coupling between surfaces. Here we have considered a thickness of three quintuple layers of $Bi_2 Se_3$ (2.6 nm) and a width of 21.0 nm. 
The size of the matrices that this structure entails will be 
$12360 \times 12360$.
Fig.~\ref{WF}(a) shows the band structure for this system, which clearly displays the presence of the Dirac point and crossing edge states inside the bulk gap. The Fermi energy is located at 0.09 eV.
In Fig.~\ref{WF}(b) we have plotted the spatial distribution of the modulus square of the wave function $|\Psi|^2$ for the two right-moving states (one of predominantly spin-up and the other of predominantly spin-down) and the two left-moving states at the 4-fold degenerate energy $E=0.1eV$. We can see that two right-moving states are edge states localized at the two external borders of the nanoribbon, while the left-moving states are interface states both localized in the region of the DW in the middle of the nanoribbon. As shown in the inset, the spin-resolved analysis of these edge states shows that the external boundaries have spin polarized states with the majority spin being opposite at the two boundaries. However, the states at the domain wall in the middle of the ribbon are non-polarized. This is in perfect agreement with the fact that each half of the ribbon is a QAHE system, where in the upper (lower) part with positive (negative) magnetization, the majority spin-up (spin-down) electrons are moving clockwise (anti-clockwise).

Therefore, if we have magnetized leads, which allow only electrons of a given spin to enter the channel, the electrons basically transport only via one edge, and by changing the direction of the current flow, the opposite spin will flow in the system through the other edge. In other words, we have a spin filter device that, by switching the direction of the voltage (or injected current) yields a spin-polarized current of opposite spin-polarization sign.

\begin{figure}[bh] \centering
\includegraphics[width=9cm,height=9cm,keepaspectratio]{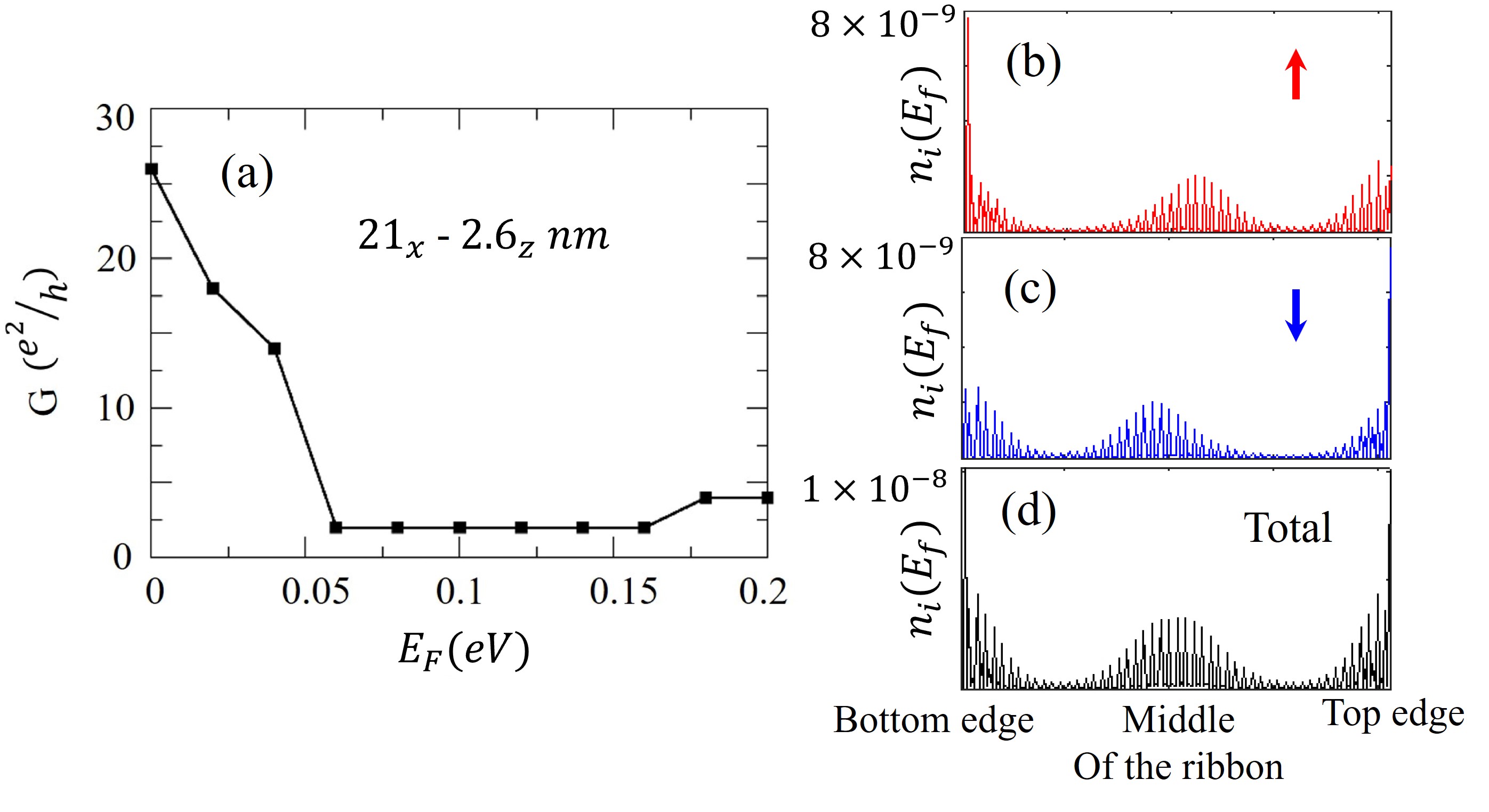}
\caption{\label{Gh} {(a) Two-terminal longitudinal conductance as a function of the Fermi energy, for the system considered in Fig~\ref{struct}(a). 
(b)-(d) Spin-resolved local density of states LDOS plotted along the width of the nanoribbon calculated when the Fermi energy is at the Dirac point. Red (panel b) and blue (panel c) colours represent spin-up and spin-down states, respectively; black (panel d) is the total LDOS.  The plots show that the bottom (top) edge of the nanoribbon is spin-polarized with spin-down (spin-up) majority electrons, while the interface states along the DM in the middle of the ribbon there are both spin-up and spin-down states, as indicated schematically in Fig.~\ref{WF}(b).  The rapid oscillations of $n_i$ in panels (b)-(d) as a function of the perpendicular coordinate $x$ come from the contribution of atoms positioned in different atomic layers (that is, having different $z$-coordinates) but with having close x-coordinate values. 
The LDOS $n_i$ is calculated using Eq.~\ref{eq:8} at zero applied voltage, namely in equilibrium, and it is in units of $\frac{e}{4\pi}$.}
}
\end{figure}

We now discuss the longitudinal conductance through the central channel, calculated using the formalism of Sec.~\ref{sec:level22}. In Fig~\ref{Gh}(a) we plot the two-terminal conductance as a function of the Fermi energy. Due to the large computational cost involved in this calculation (the Green's function matrices that we have to invert become exceptionally large), we have calculated the conductance for only a few values of the Fermi energy in the relevant energy window $0 \le E_{\rm F} \le 0.2$ eV dictated by the band structure of  Fig.~\ref{WF}(a) in order to ascertain its general behavior.  
We can clearly distinguish three different regimes. In the first one,
$0 \le E_{\rm F} \le 0.06$ eV, the Fermi energy in inside the continuum of the bulk states of the valence bands. The conductance is large and a monotonically decreasing function of the energy.
The second regime, when the Fermi energy varies within the energy interval $[0.06, 0.16]$eV which corresponds to the exchange gap. Here we only have the four quasi-degenerate chiral states (two right-moving edge states and two left-moving interface states). Therefore the two-terminal longitudinal conductance is quantized and equal to  
$G=\frac{2e^2}{h}=2 G_0$ corresponding to the contribution of the dissipationless current carried by two edge states. In a four-terminal system, this situation would give rise to a QHAE with Hall resistance equal to $R_{\rm H} = \frac{h}{2 e^2}$.  Interestingly, 
when the current is injected from the left lead into the right lead, the chiral nature of the 1D conducting boundary states implies that transport takes place along the spin-polarized edges of the ribbon.
Viceversa, when the current is injected from the right lead, it flows dissipationlessly into the right lead via the unpolarized interface states along the DW in the middle of ribbon.
Finally, the third regime corresponds to the energy interval 
$E_{\rm F}> 0.16$ eV, where the Fermi energy starts to probe states above the exchange gap belonging to the conduction band. In contrast to the first regime, where $E_{\rm F}$ is below the exchange gap inside the bulk valence band continuum, here the states of the bottom of the conduction band are discrete. Clearly this is in part caused by the finite width of the nanoribbon. As shown in Ref.~\cite{pertsova2016}, these states are non-chiral and appear outside the exchange gap. Just like in a quantum point contact, when by increasing the Fermi energy, additional discrete 1D states get occupied and contribute to the transport, each increasing the conductance by $G_0$, here these non-chiral states cause the quantized step-wise increase of the conductance. However, in contrast to the in-gap chiral states, these additional contribution are not robust against disorder and are going to be back-scattered by impurities and imperfections. Note that deviations from perfect quantization of the QAH resistance in magnetic TIs, signalled by a non-zero longitudinal resistance, have recently been discussed in Ref. ~\cite{Rosen2022}, where it was pointed out that, in contrast to what is generally assumed, current can flow not only via chiral edge states but also through dissipative 2D bulk states.

In Fig~\ref{Gh}(b) we plot the spin-resolved and the total local density of states $n_i^\sigma (E_{\rm F})$ along the nanoribbon width, calculated using Eq.\ref{eq:9} at equilibrium, that is at zero applied voltage, when $E_{\rm F}$ is at the Dirac point.
This figure shows that the states inside the exchange gap displayed in Fig.~\ref{WF}(a)  are indeed spin-polarized edge and interface states localized respectively along the top and bottom nanoribbon edges as well as in the middle of the nanoribbon where the DW is located. This figure is therefore completely consistent with the results of the longitudinal conductance shown in Fig.~\ref{Gh}(a).

\subsection{\label{sec:level31} Perpendicular domain wall}
Next, we consider the geometry shown in Fig~\ref{struct}(b), 
The DW, placed in the middle of central channel, is now orthogonal to the nanoribbon length and to the direction of the longitudinal transport, and it is going to act as a spin-dependent potential barrier that will affect the two terminal longitudinal conductance. 

In this case, translational invariance along the longitudinal (y) direction is broken, and therefore we cannot plot the 1D band structure as in the previous case.
Nevertheless, we can gain an understanding of the electronic structure of the system by plotting the local density density of states (LDOS) close to the left and right leads. For this purpose, 
we make use of the iterative Green's function formalism also employed to analyze transport. As shown in Eq.~\ref{eq:9}, at zero applied voltage the lesser Green's function yields precisely the {\it equilibrium} LDOS at a given position at energy $E$.

To compute the LDOS at a specific space position of the system, we essentially address different parts of the nanoribbon by introducing a sequence of different layers that make up the central Hamiltonian $H_C$ in Eq~\ref{eq:5}. Each layer of $H_C$ is a unit cell inside the channel, which is repeated along its length. In contrast to the previous case of longitudinal DW where iterative layers defining $H_C$ are all the same, now for this nanoribbon configuration $H_C$ will change when the iterative process reaches the DW.

\begin{figure}[h] \centering
\includegraphics[width=9cm,height=9cm,keepaspectratio]{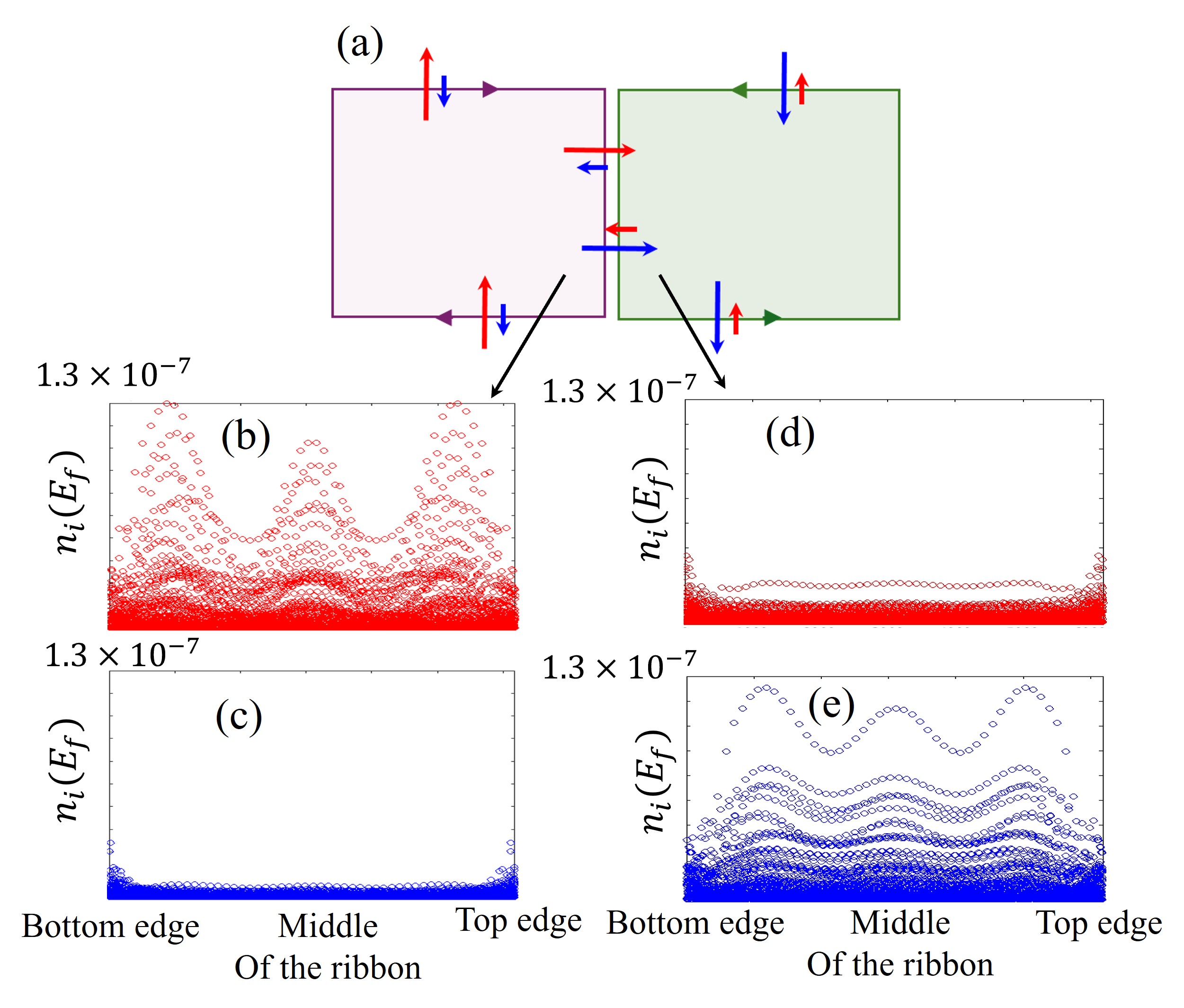}
\caption{\label{dp} {(a) Schematic figure of the edge states in the presence of a perpendicular DW. (b)-(e) Spin-resolved LDOS calculated at $E_{\rm F}= 0.1$ eV, plotted along the width of the nanoribbon just on the left (b-c, positive magnetization) and on the right (d-e, negative magnetization) of the DW. Red and blue colours represent up and down spins, respectively. The rapid oscillations of $n_i$ in panels (b)-(e) as a function of the perpendicular coordinate $x$ come from the contribution of atoms positioned in different atomic layers (that is, having different $z$-coordinates) but having close x-coordinate values. The LDOS $n_i$ is in units of $\frac{e}{4\pi}$.} 
}
\end{figure}

In Fig~\ref{dp} we plot the LDOS along the DW direction at a very close distance from the DW itself, when $E_{\rm F} = 0.01$ eV. Note that this is essentially the energy at the Dirac point of the edge states in the band structure for the parallel DW case shown shown in Fig.~\ref{WF}, located in the middle of the exchange gap. In principle, it is not clear that this energy should be at all relevant for the present case. However, not having other reference points, it is useful to consider it. Here panels (b)-(e) represent the LDOS plotted along the width of the nanoribbon just
on the left (b and c, positive magnetization) and on the right (d and
e, negative magnetization) of the DW. Red and blue curves refer again to spin-up and spin-down states respectively. The figures exhibits two main features. The first one is the opposite character of the strong spin polarization on the left and on the right of the DW, where majority and minority spins are interchanged. The second feature is the presence of a non-zero LDOS along all the width of nanoribbon for both spin-up and spin-down, suggesting the existence of 1D chiral edge and interface states as indicated schematically in panel (a). The rapid oscillations of $n_i$ in panels (b) and (e) as a function of the perpendicular coordinate $x$ come from the contribution of atoms positioned in different atomic layers (that is, having different $z$-coordinates) but having close x-coordinate values.
Physically they might be caused by backscattaring at the sharp corners where edge and interface states join.
 
\begin{figure}[h] \centering
\includegraphics[width=7cm,height=6cm,keepaspectratio]{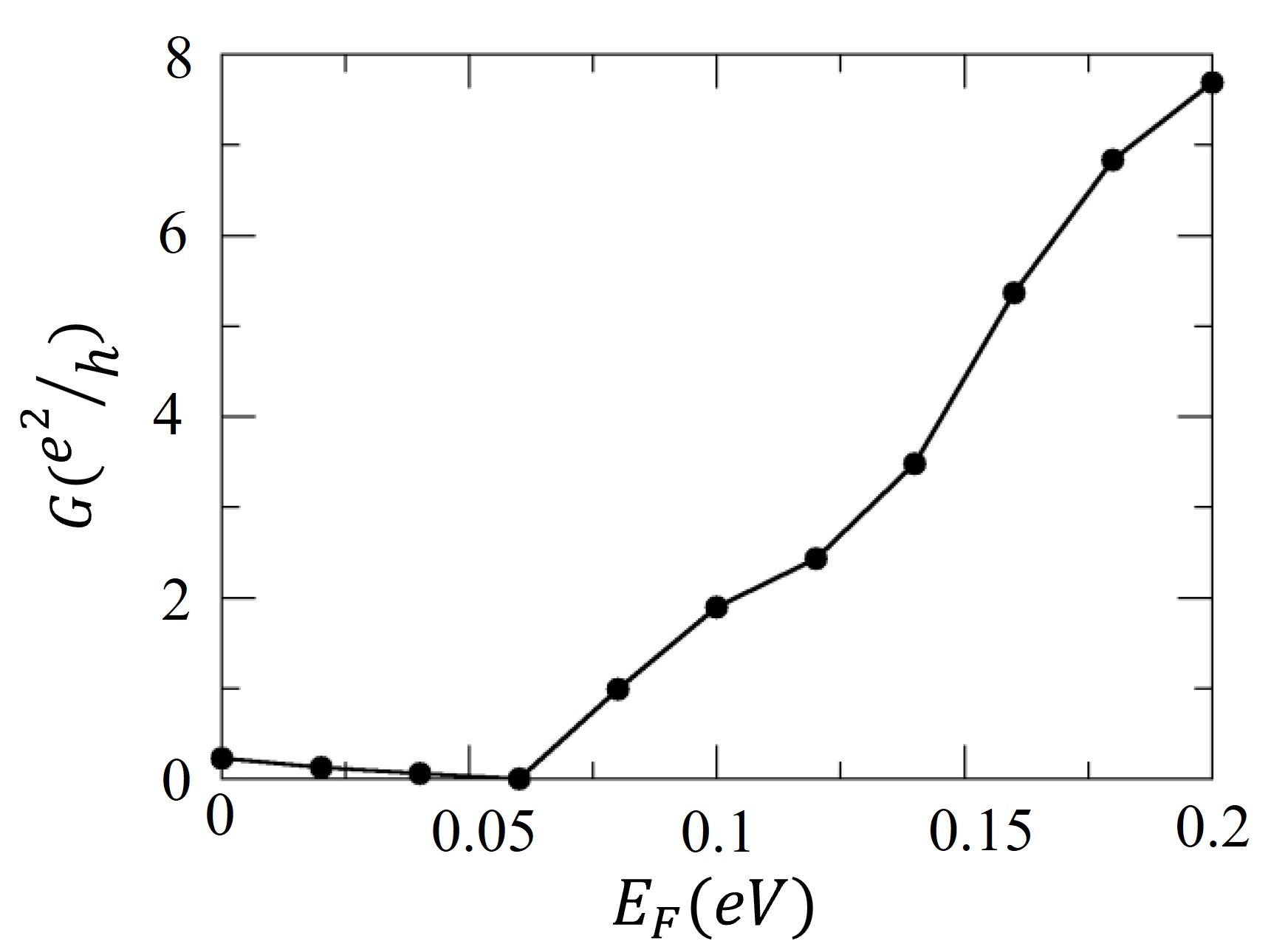}
\caption{\label{Gt} {Two-terminal longitudinal conductance plotted as a function of the Fermi energy for the system considered in Fig~\ref{struct}(b) with width 21 nm and three-quintuple-layers nanoribbon (corresponding to a thickness of 2.6 nm).
}}
\end{figure}

We now analyze the two-terminal conductance for this geometry as a function of $E_{\rm F}$, as shown in Fig.~\ref{Gt}. Two main energy regimes stand out: (i) when $E_{\rm F} < 0.06$ eV the conductance is very small, $G << G_0e$; for $E_{\rm F} > 0.06$ eV, the conductance grows approximately linearly with $E_{\rm F}$. In order to elucidate the mechanism of this surprising behavior, we first note that the value $E_{\rm F} = 0.06$ eV signals the beginning of the exchange gap for the parallel DW case, hosting chiral edge and interface states, and it is exactly the lower limit of the energy interval where the conductance in that case is quantized. Again, although we do not have direct strong physical arguments supporting the assumption that this energy should also be relevant for the present case, our intuition suggests that this occurrence is not a sheer coincidence. We have already observed that, on the two sides of the DW and very close to it, 
the LDOS at $E_{\rm F} = 0.1$ eV (the energy at the Dirac point for the parallel DW case) is consistent with the one of chiral spin polarized interface states joining the corresponding edge states which propagate on the two nanoribbon external edges, see Fig.~\ref{dp}.
We pursue this trail by computing the LDOS at the two ends of the central channel, far away from the DW for a few different energies. We choose these value guided by the band structure of Fig.~\ref{WF}, which now should be somewhat relevant for this case given the large distance from the DW. The panels on the left (a-c) show the LDOS calculated at the position the first unit cell in the channel, which is closest to the left lead, while in the panels on the right (d-f) LDOS is calculated  for the last unit cell in the channel, which is closest to the right lead. 
At $E= 0.002$, (panels a and d) the LDOS shows the expected strong spin-polarization; however it is pretty much uniform along the width of the nanoribbon. Given the fact that this energy is located deep in the valence bands of Fig.~\ref{WF}(a), we interpret this LDOS as the one pertaining to bulk, non chiral states spread throughout the nanoribbon. When $E_{\rm F}$ is in this region, it is primarily these states which would contribute to the longitudinal conductance ~\cite{Rosen2022}. However, when the injected current from the left lead carried by these states reaches the DW, finds on the other side a LDOS with where only states with the opposite spin sign are available and cannot propagate further. Hence the conductance is very small. In way this is the same mechanism responsible for the giant magneto-resistance in magnetic multilayers. 
When  $E= 0.002$ eV, (panels c and f) we expect LDOS to still correspond to non chiral bulk states, since this value of the energy is located way-up in the conduction band of the band structure of Fig.~\ref{WF}(a). However, now the magnetic properties are quite different. Indeed the spin polarization on both sides of the DW is much smaller. In this case, a current injected from the left lead at this value of $E_{\rm F}$ and should the able to flow (unhindered in the absence of imperfections) and be collected at the right lead. The conductance should be proportional to the cumulative contribution of all the occupied states up to $E_{\rm F}$, as found in the calculation. 
The intermediate energy region between these two values, represented by panels b and, is the the most interesting one, since $E$ is now inside the exchange gap and therefore 1D chiral edge and interface states are the only ones active. We can imagine that when the current is injected from the left lead, it flows first along the upper edge state. At the DW is continues along the chiral interface state moving down until it reaches the lower edge state when it can continue to the right until it is collected at the right lead. Now the opposite spin polarization in the two halves of the nanoribbons does not prevent current flow completely. Two possible reasons for this are: (i) some minority spin states of the the minority spin contribution in the lower edge (having therefore the same ``correct´´ spin of the chiral states originating from the upper edge) states might be larger; (ii) while majority up-spin electrons travel downward along the DW interface state might spin-flip due to the presence of spin-orbit interaction.  
However, when the Fermi energy is in this energy window transport in general will not be quantized, although it amusing to note that the longitudinal conductance is $G \approx 2\; G_0$..

\begin{figure*}[t] \centering
\includegraphics[width=12cm,height=12cm,keepaspectratio]{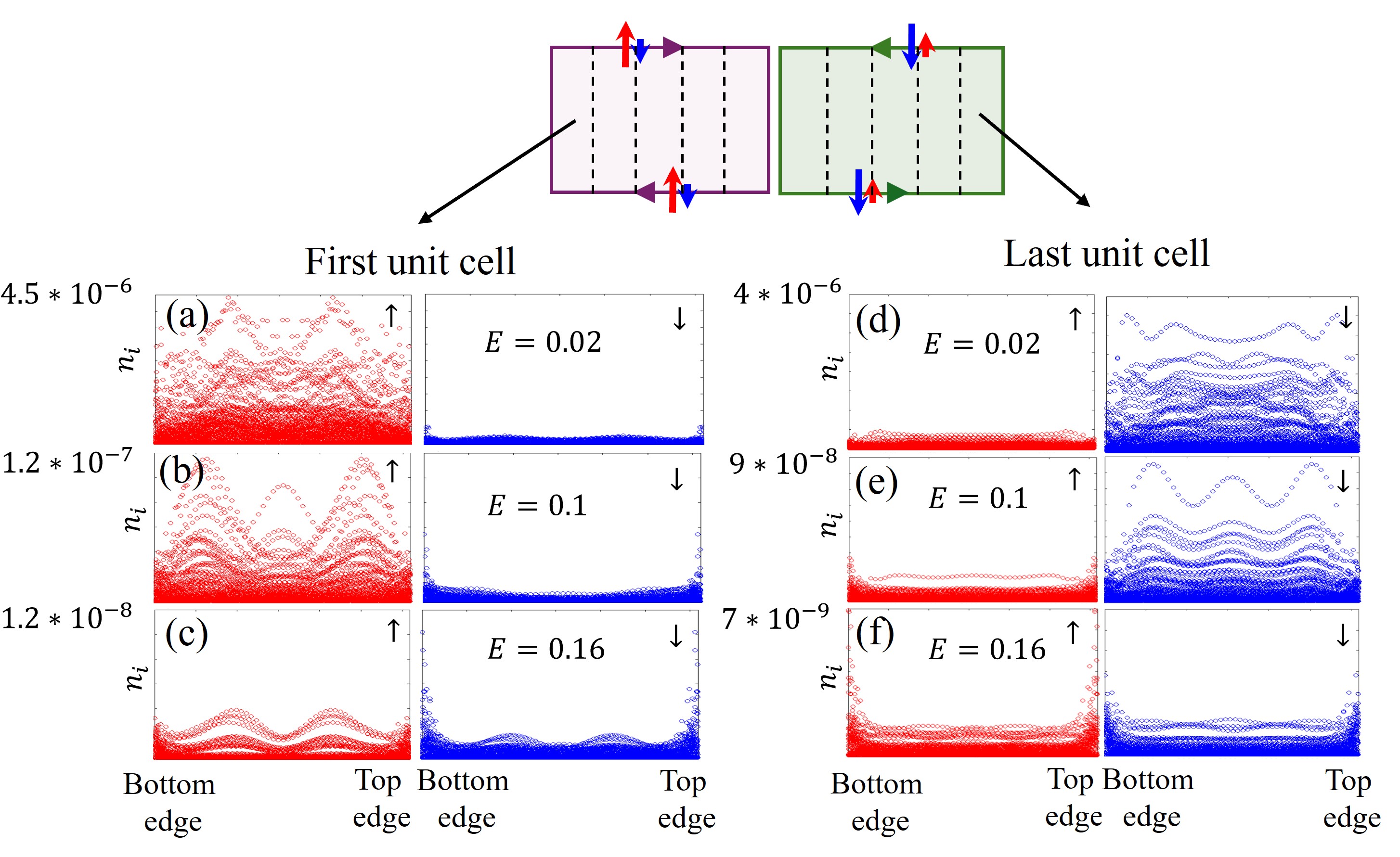}
\caption{\label{dpe} {Spin-resolved LDOS ($n_i$) at two positions along the length of the nanoribbon at different energies $E$. (a)-(c) $n_i$ calculated at the position the first unit cell in the channel which is closest to the left lead. (d)-(f) $n_i$ calculated  for the last unit cell in the channel which is closest to the right lead. Red and blue colours represent up and down spins, respectively. The LDOS $n_i$ is in units of $\frac{e}{4\pi}$. Energy units are in eV.}
}
\end{figure*}

\section{Conclusion}
Using an atomistic tight-binding model that captures realistically the microscopic features of magnetic TI heterostructures, we have analyzed the charge and spin characteristics of the 1D chiral edge and interface states in a magnetized $Bi_2Se_3$ TI nanoribbon in the presence of one magnetic domain wall (DW) of two possible different kinds 

For a DW parallel to the longitudinal direction of the nanoribbon, exact quantization of the anomalous Hall conductance is guaranteed provided that (i) the Fermi energy is located inside the magnetic exchange energy gap where only chiral edge and interface states are present (ii)  the width of the nanoribbon is large enough to prevent the coupling of opposite channels. The edge states at the  external boundaries of the system are spin-polarized and can be tuned by means of an external voltage. In the case of perpendicular domain wall, back-scattering mechanisms caused by the DW give rise to a deviation from the perfect quantization. In particular, the longitudinal conductance vanishes identically when the Fermi energy is smaller than a critical value identified as the lower edge of the exchange gap in the 1D band structure of the parallel case, and increases linearly afterward. A more detailed understanding of these results could perhaps be achieved by examining the  space dependence of the steady-state spin-polarized current density ~\cite{Pournaghavi2018}. 

Further application of the approach presented here could yield valuable insights into the impact of more intricate spin textures, for example topological spin excitations such as skyrmions, on both charge and spin conductance. The size of these skyrmions is particularly crucial for a detailed analytical and computational analysis of the electronic structure and especially the resulting quantum transport. Recent studies of TI/magnetic-layer heterostructures have reported evidence of the existence of skyrmions of relatively small size of the order of 15 nm at room temperature ~\cite{Liu2023} on the surface of a topological insulator which could be analyzed with our microscopic approach.
Outstanding issues, presently intensively investigated, include the use of the unique spin properties of the conducting TI gapless surface and edge states to manipulate skyrmions\cite{Hurst2015, Chen2019, Kurebayashi2022}. 
Similarly, spin-orbit torque, thought to be particularly large in TI thin films\cite{Mellnik2014, ndiaye17:014408, manchon2019}, 
can be used to affect and possibly revert the magnetic properties of these topological spin structures.

\section{Acknowledgments} This work was supported by the Faculty of Technology and by the Department of Physics and Electrical Engineering at Linnaeus University (Sweden).  We acknowledge financial support from the Swedish Research Council (VR 2021-046229) and Carl Tryggers Stiftelsen (CTS 20:71). 
The computations at Linnaeus University
were enabled by resources provided by the National Academic
Infrastructure for Supercomputing in Sweden (NAISS) at Dardel,
partially funded by the Swedish Research Council through grant
agreement no. 2022-06725.
\bibliography{Domain}
\end{document}